\documentclass[12pt]{book}

\usepackage{plenum}

\begin{document}

\title{IS THE COSMOLOGICAL CONSTANT NON-ZERO?}
\author{Paul H. Frampton}
\affiliation{Department of Physics and Astronomy, \\
University of North Carolina,\\
Chapel Hill, NC 27599-3255.}

\section{THE ISSUE OF $\Lambda$}

The present talk is similar to one at COSMO-98 last month.
As is well explained in standard
reviews of the cosmological constant
[\cite{weinberg, ng}] the theoretical expectation
for $\Lambda$ exceeds its observational value
by $120$ orders of magnitude.

In 1917, Einstein looked for a static solution of
general relativity for cosmology and added a new
$\lambda$ term:

\begin{equation}
R_{\mu\nu} - \frac{1}{2}g_{\mu\nu}R - \lambda g_{\mu\nu}
= - 8 \pi G T_{\mu\nu}
\end{equation}
A $\lambda > 0$ solution exists with $\rho = \frac{\lambda}{8 \pi G}$,
radius $r(S_3) = (8 \pi \rho G)^{-1/2}$ 
and mass $M = 2 \pi^2 r^3 \rho = \frac{\pi}{4 \sqrt{\lambda} G}$.  

In the 1920's the universe's expansion became 
known (more red shifts than blue shifts). In 1929, 
Hubble enunciated his law that recession velocity 
is proportional to distance.

Meanwhile Friedmann (1922) discovered the now-standard
non-static model with metric:
\begin{equation}
ds^2 = dt^2 - R^2(t) \left[ \frac{dr^2}{1 - kr^2} +
r^2(d \theta^2 + sin^2\theta d \phi^2) \right]
\end{equation}

In 1923, Einstein realized the dilemna. He wrote to his friend Weyl:\\
``If there is no quasi-static world, then away with the cosmological 
term''.

Setting $\Lambda = 0$ does not increase symmetry. In fact, the
issue is one of {\it vacuum energy} density as follows:

In vacuum:
\begin{equation}
<T_{\mu\nu} = -<\rho> g_{\mu\nu}
\end{equation}
which changes the $\lambda_{eff}$ by:
\begin{equation}
\lambda_{eff} = \lambda + 8 \pi G <\rho>
\end{equation}
or equivalently:
\begin{equation}
\rho_V = <\rho> + \frac{\lambda}{8 \pi G} = \frac{\lambda_{eff}}
{8 \pi G}
\end{equation}
The observational upper limit on $\lambda$ comes from:
\begin{equation}
\left( \frac{dR}{dt} \right)^2 = -k + \frac{1}{3} R^2 
(8 \pi G \rho + \lambda)
\end{equation}
which expresses conservation of energy and leads to the 
upper bound $|\lambda_{eff}| \leq H_0^2$.

This translates into $|\rho_V| \le 10^{-29} g/cm^3$.
In high-energy units we use $1g \sim 10^{33}eV$
and $(1 cm)^{-1} \sim 10^{-4}$eV to rewrite $|\rho_V| \le 
 [(1/100)eV]^4$

A ``natural'' value in quantum gravity is:
\begin{equation}
|\rho_V| = (M_{Planck})^4 = (10^{28} eV)^4
\end{equation}
which is $10^{120}$ times too big. This has been called the
biggest error ever made in theoretical physics!

Even absent the $(M_{Planck})^4$ term field theory with
spontaneous symmetry breaking leads one
to expect $<\rho>~~~\gg [(1/100)eV]^4$.
As examples, QCD confinement suggests $<\rho> \sim (200MeV)^4$,
which is $10^{40}$ times too big and electroweak spontaneous
symmetry breaking would lead to $<\rho> \sim (250GeV)^4$
which is $10^{52}$ times too big.
This is the theoretical issue. I will briefly
mention four approaches to its solution.

\subsection{(1) Supersymmetry, Supergravity, Superstrings.}

According to global supersymmetry:
\begin{equation}
\{Q_{\alpha}, Q_{\beta}^{\dagger}\}_{+} = (\sigma_{\mu})_{\alpha\beta} P^{\mu}
\end{equation}
and with unbroken supersymmetry:
\begin{equation}
Q_{\alpha} | 0 > = Q_{\beta}^{\dagger} | 0 > = 0
\end{equation}
which implies a vanishing vacuum value for  $<P_{\mu}>$ and hence zero 
vacuum energy as required for vanishing $\Lambda$.

With global supersymmetry promoted to local
supersymmetry the expression for the potential is more
complicated than this (one can even have $V < 0$).

When supersymmetry is broken, however, at $\geq 1$ TeV one expects again
that $|\rho_V| > (1 TeV)^4$ which is $10^{54}$
times too big.

So although unbroken supersymmetry looks
highly suggestive, broken supersymmetry does
not help. The same is generally true for
superstrings.

One new and exciting approach - still in its infancy -
involves the compactification of the Type IIB superstring
on a manifold $ S^5 \times AdS_5$ and give rise
to a 4-dimensional ${\cal N} = 4$ SU(N) supersymmetric
Yang-Mills theory, known to be conformal.
Replacing $S^5$ 
by an orbifold 
$S^5 / \Gamma$ 
can lead to
${\cal N} = 0$ non-supersymmetric
SU(N) gauge
theory and probably (this is presently 
being checked; see {\it e.g.} [\cite{frampton}])
retain conformal symmetry. If so one may achieve $<\rho> = 0$
without supersymmetry.

\subsection{(2) Quantum Cosmology.}

The use of wormholes to derive $\Lambda \rightarrow 0$
has been discredited because of (a) the questionable use of
Euclidean gravity, (b) wormholes, if they exist, become
macroscopically large and closely-packed, at variance with
observation.

\subsection{(3) Changed Gravity.}

An example of changing gravity theory [\cite{FNV}] is to make
$g = det g_{\mu\nu}$ non-dynamical
in the generalized action:
\begin{equation}
S = -\frac{1}{16 \pi G} \int dx [R + L (g -1) ]
\end{equation}
where L is a Lagrange multiplier. One then finds by
variation that $R = -4\Lambda = constant$. Minimizing the action
gives $\Lambda = 2\sqrt{6} \pi / \sqrt{V}$ where
V is the spacetime volume.

In the path integral
\begin{equation}
Z = \int d\mu(\Lambda) exp (3 \pi / G \Lambda)
\end{equation}
the value $\Lambda \rightarrow 0^+$ is
exponentially favored.
 
\subsection{(4) The Anthropic Principle.}

If $\Omega_{\Lambda} \gg 1$ rapid exponential expansion
prohibits gravitational condensation to clumps of matter.
This requires $\Omega_{\Lambda} < 400$.

On the other hand if $\Omega_{\Lambda} \ll 0$ the universe
collapses at a finite time, and there is not
enough time for life to evolve.
For example, if $\Lambda = - (M_{Planck})^4$, R reaches only 0.1mm
($10^{-30}$ of its present value). Taken together
these two considerations lead to
\begin{equation}
-1 \le \Omega_{\Lambda} \le 400
\end{equation} 
-- quite a strong constraint. 

This shows how important it is to life that $\Lambda$ is 
very much closer to zero than
to $(M_{Planck})^4$ or even $E^4$ where
$E$ is
any vacuum energy scale familiar to High Energy physics.

\section{CBR TEMPERATURE ANISOTROPY}

The Cosmic Background Radiation (CBR) was discovered [\cite{PW}]
in 1965 by Penzias and Wilson.
But detection of its temperature anisotropy waited until
1992 when [\cite{S1,S2}] the Cosmic Background Explorer (COBE) satellite
provided impressive experimental support for the Big
Bang model. COBE
results are consistent with a scale-invariant spectrum
of primordial scalar density fluctuations, such as might 
be generated by quantum fluctuations during inflation
[\cite{BST,S,GP,H,G,L,AS}].
COBE's success inspired many further experiments with
higher angular sensitivity than COBE ($\sim 1^o$).

NASA has approved a satellite mission MAP (Microwave
Anisotropy Probe)
for 2000. ESA has approved the Planck surveyor -
even more accurate than MAP - a few years later in 2005.

With these experiments, the location of the first
accoustic (Doppler) peak and possible subsequent
peaks will be resolved.

The Hot Big Bang model is supported
by at least three major triumphs:
\begin{itemize}
\item the expansion of the universe
\item the cosmic background radiation
\item nucleosynthesis calculations
\end{itemize}
It leaves unanswered two major questions:
\begin{itemize}
\item the horizon problem
\item the flatness problem
\end{itemize}

{\it The horizon problem.}
When the CBR last scattered, the age of the universe was
$\sim 100,000 y$. The horizon size at that recombination time subtends
now an angle $\sim \pi/200$ radians. On the celestial sphere
there are 40,000 regions never causally-connected in the unadorned
Big Bang model. Yet their CBR temperature is the same to one
part in $10^5$ - how is this uniformity arranged?

{\it The flatness problem.}
From the equation (for $\Lambda = 0$)
\begin{equation}
\frac{k}{R^2} = (\Omega - 1) \frac{\dot{R}^2}{R^2}
\end{equation}
and evaluate for time $t$ and the present $t - t_0$, using
$R \sim \sqrt{t} \sim T^{-1}$:
\begin{equation}
(\Omega_t - 1) = 4H_0^2 t^2 \frac{T^2}{T_0^2} (\Omega_0 - 1)
\label{Omega}
\end{equation}
Now for high densities:
\begin{equation}
\frac{\dot{R}^2}{R^2} = \frac{8 \pi G \rho}{3} \simeq \frac{8 \pi G g a T^4}{6}
\end{equation}
where $a$ is the radiation constant $= 7.56 \times 10^{-9} erg ~~~m^{-3} ~~~K^{-4}$.

From this we find 
\begin{equation}
t (seconds) = (2.42 \times 10^{-6}) g^{-1/2} T(GeV)^{-2}
\end{equation}
and thence by substitution in Eq. (\ref{Omega})
\begin{equation}
(\Omega_t - 1) = (3.64 \times 10^{-21}) h_0^2 g^{-1} T(GeV)^{-2} (\Omega_0 - 1)
\end{equation}
This means that if we take, for example, $t = 1 second$ when $T \simeq 1$ MeV,
then $|\Omega_t - 1|$ must be $ < 10^{-14}$ for $\Omega_0$ to be of order unity
as it is now. If we go to earlier cosmic time, the fine tuning of
$\Omega_t$ becomes even stronger if we want the present universe to
be compatible with observation. Why then is $\Omega_t$ so extremely
close to $\Omega_t = 1$ in the early universe?	

{\it Inflation}
Both the horizon and flatness problems can be solved in the 
inflationary scenario which has the further prediction (in general)
of flatness. That is, if $\Lambda = 0$:
\begin{equation}
\Omega_m = 1
\label{flatflat}
\end{equation}
or, in the case of $\Lambda \neq 0$ (which is allowed by inflation):
\begin{equation}
\Omega_m + \Omega_{\Lambda} = 1
\label{flat}
\end{equation}
We shall see to what extent this prediction, Eq.(\ref{flat}), is
consistent with the present observations.

\bigskip
\bigskip

The goal of the CBR experiments [\cite{davis,bond,steinhardt,loeb}]
is to measure the temperature autocorrelation function.
The fractional perturbation as a function of direction $\underline{\hat{\bf n}}$
is expanded in spherical harmonics:
\begin{equation}
\frac{\Delta T({\underline{\hat{\bf n}})}}{T} = \sum_{lm} a_{lm} Y_{lm} (\underline{\hat{\bf n}})
\end{equation}
The statistical isotropy and homogeneity ensure that
\begin{equation}
< a^{\dagger}_{lm} a_{l'm'} > = C_l \delta_{ll'} \delta_{mm'}
\end{equation}
A plot of $C_l$ versus $l$ will reflect oscillations in the baryon-photon
fluid at the surface of last scatter. The first Doppler, or accoustic,
peak should be at $l_1 = \pi / \Delta \Theta$ where $\Delta \Theta$ is the angle
now subtended by the horizon at the time of last scatter: the recombination
time at a red-shift of $Z \simeq 1,100$.

\subsection{The special case $\Lambda = 0$}

When $\Lambda = 0$, the Einstein-Friedmann cosmological equation
can be solved analytically (not generally true if $\Lambda \neq 0$). 
We will find $l_1 \sim 1/\sqrt{\Omega_m}$ as follows. Take:
\begin{equation}
ds^2 = dt^2 - R^2 \left[ d\Psi^2 + sinh^2~~\Psi~~d\Theta^2
+ sinh^2~~\Psi~~sin^2~~\Theta ~~d\Phi^2 \right]
\label{metric}
\end{equation}
For a geodesic $ds^2 = 0$ and so:
\begin{equation}
\frac{d \Psi}{d R} = \frac{1}{R}
\end{equation}
The Einstein equation is
\begin{equation}
\left( \frac{\dot{R}}{R} \right)^2 = \frac{8 \pi G \rho}{3} + \frac{1}{R^2}
\end{equation}
so that
\begin{equation}
\dot{R}^2 R^2 = R^2 + a R
\end{equation}
with $a = \Omega_0 H_0^2 R_0^3$ and hence
\begin{equation}
\frac{d \Psi}{d R} = \frac{1}{\sqrt{ R^2 + a R}}
\end{equation}
This can be integrated to find
\begin{equation}
\Psi_t = \int_{R_t}^{R_0} \frac{dR}{\sqrt{(R +a/2)^2 - (a/2)^2}}
\end{equation}
The substitution $R = \frac{1}{2} a (coshV - 1)$ leads to
\begin{equation}
\Psi_t = cosh^{-1}(\frac{2R_0}{a} - 1) - cosh^{-1}(\frac{2R_t}{a} - 1)
\end{equation}
Using $sinh(cosh^{-1}x) = \sqrt{x^2 - 1}$ gives
\begin{equation}
sinh \Psi_t = 
\sqrt{\left( \frac{2 (1 - \Omega_0)}{\Omega_0} + 1 \right)^2 - 1}
- \sqrt{\left( \frac{2 (1 - \Omega_0) R_t}{\Omega_0 R_0} + 1 \right)^2 - 1}
\label{Psit}
\end{equation}
The second term of Eq.(\ref{Psit}) is negligible as $R_t/R_0 \rightarrow 0$ 
With the metric of Eq.(\ref{metric}) the angle subtended now by the horizon
then is
\begin{equation}
\Delta \Theta = \frac{1}{H_tR_t sinh \Psi_t}
\end{equation}
For $Z_t = 1,100$, the red-shift of recombination
one thus finds
\begin{equation}
l_1 (\Lambda = 0) \simeq \frac{2 \pi Z_t^1/2}{\sqrt{\Omega_m}} \simeq \frac{208.4}{\sqrt{\Omega_m}}
\end{equation}
This is plotted in Fig. 1 of [\cite{FNR}].

\subsection{The general case $\Lambda \neq 0$}

When $\Lambda \neq 0$
\begin{equation}
\dot{R}^2 R^2 = -kR^2 + aR +\Lambda R^4/3
\end{equation}
It is useful to define the contributions to the
energy density $\Omega_m = 8 \pi G \rho/3H_0^2$, $\Omega_{\Lambda}
= \Lambda/3H_0^2$, and $\Lambda_C = -k/H_0^2R_0^2$. These satisfy
\begin{equation}
\Omega_m + \Omega_{\Lambda} + \Omega_C = 1
\end{equation}
Then
\begin{equation}
l_1 = \pi H_t R_t sinh \Psi_t
\end{equation}
where
\begin{equation}
\Psi_t = 
\sqrt{\Omega_C} \int_1^{\infty} \frac{d w}{\sqrt{\Omega_{\Lambda} +
\Omega_C w^2 +\Omega_m w^3}}
\end{equation}
After changes of variable one arrives at
\begin{equation}
l_1 = \pi \sqrt{\frac{\Omega_0}{\Omega_C}} \sqrt{\frac{R_0}{R_t}}
sinh \left(
\sqrt{\Omega_C} \int_1^{\infty} \frac{d w}{\sqrt{\Omega_{\Lambda} +
\Omega_C w^2 +\Omega_m w^3}}
\right)
\label{l1}
\end{equation}
(For positive curvature ($k = +1$) replace $sinh$ by $sin$).
For the case $k = 0$, the flat universe predicted by inflation,
with $\Omega_C = 0$ Eq.(\ref{l1}) reduces to
\begin{equation}
l_1 = \pi \sqrt{\Omega_m} \sqrt{\frac{R_0}{R_t}}
 \int_1^{\infty} \frac{d w}{\sqrt{\Omega_{\Lambda} +
 +\Omega_m w^3}}
\end{equation}
These are elliptic integrals, easily do-able by Mathematica. 
They resemble the formula for the age of the universe:
\begin{equation}
A = \frac{1}{H_0}  \int_1^{\infty} \frac{d w}{w \sqrt{\Omega_{\Lambda} +
\Omega_C w^2 +\Omega_m w^3}}
\end{equation}
In Fig. 2 of [\cite{FNR}] there is a plot of $l_1$ versus $\Omega_m$
for $\Omega_C = 0$. In Fig. 3 are the main result of the iso-$l_1$
lines on a $\Omega_m-\Omega_{\Lambda}$ plot for general $\Omega_C$
with values of $l_1$ between 150 and 270 in increments
$\Delta l_1 = 10$. The final Fig. 4 of [\cite{FNR}] gives a
three-dimensional plot of $\Omega_m - \Omega_{\Lambda} - l_1$.

\bigskip
\bigskip

We can look at the cumulative world data on $C_l$ versus $l$.
Actually even the existence of the first
Doppler peak is not certain but one can see evidence for the
rise and the fall of $C_l$. 
In Fig. 2 of [\cite{K}] we see such 1998 data and with some licence say that
$150 \leq l_1 \leq 270$. 

The exciting point is that the data are expected to improve
markedly in the next decade. In Fig. 3 of [\cite{K}] there
is
an artist's impression of both MAP data (expected 2000) and Planck
data(2006); the former should pin down $l_1$ with a small error
and the latter is expected to give accurate
values of $C_l$ out to $l = 1000$.

\bigskip
\bigskip

But even the spectacular accuracy of MAP and Planck will
specify only one iso-$l_1$ line in the $\Omega_m - \Omega_{\Lambda}$
plot and not allow unambiguous determination of $\Omega_{\Lambda}$.

\bigskip

Fortunately this ambiguity can be removed by a completely independent
set of observations.

\section{III. HIGH-Z SUPERNOVAE IA.}

In recent years several supernovae (type IA) have been discovered with high
red-shifts $Z > 0.3$ (at least 50 of them). An example of a high red-shift
is $Z = 0.83$. How far away is that in cosmic time?
For matter-domination
\begin{equation}
\left( \frac{R_0}{R_t} \right) = \left( \frac{t_0}{t} \right)^{2/3} = (Z +1)
\end{equation}
so the answer is $t = t_0/2.83$. For $t_0 = 14Gy$ this implies $t \simeq 6Gy$.
Thus this supervova is older than our Solar System and
the distance is over half way back to the Big Bang!

These supernovae were discovered [\cite{SN1,SN2}]
by a 4m telescope then their light-curve
monitored by the 10m telescope KEK-II on Mauna Kea, Hawaii 
{\it and/or} the Hubble Space Telescope. 
The light curve is key, because study
of nearby supernovae suggests that the breadth of the light curve
{\it i.e.} the fall in luminosity in 15 days following its peak
is an excellent indicator of absolute luminosity. Broader (slower)
light curves imply brighter luminosity. Clever techniques compare
the SN light-curve to a standard template.

It is worth pointing out that although 
these SN are very far away - over 50\% back to the 
Big Bang they do not penetrate as far back as the CBR discussed
earlier which goes 99.998\% back to the Big Bang (300,000y out
of 14,000,000,000y).

Because of the high $Z$, just one of these observations, and certainly
50 or more of them, have great influence on the estimation
of the deceleration parameter $q_0$ defined by
\begin{equation}
q_0 = -\frac{\mbox {\"{R}}R}{\dot{R}^2}
\end{equation}
which characterizes departure from the linear Hubble relation $Z = \frac{1}{H_0}d$.
In the simplest cosmology ($\Lambda = 0$) one
expects that $q_0 = + 1/2$, corresponding to a {\it deceleration} in
the expansion rate. 

\bigskip
\bigskip

The startling result of the high-Z supernovae observations is that
the deceleration parameter comes out {\it negative} $q_o \simeq
 -1/2$ implying
an {\it accelerating} expansion rate.

Now if the only sources of vacuum energy driving the expansion
are $\Omega_m$ and $\Omega_{\Lambda}$ there is the relationship
\begin{equation}
q_0 = \frac{1}{2} \Omega_m - \Omega_{\Lambda}
\label{q0}
\end{equation}
\bigskip
\bigskip

So we add a line on the $\Omega_m - \Omega_{\Lambda}$ plot
corresponding to Eq.(\ref{q0}) with $q_0 = -1/2$. Such a line is
orthogonal to the iso-$l_1$ lines from the CBR 
Doppler peak and the intersection gives the
result that values $\Omega_m \simeq 0.3$ and $\Omega_{\Lambda} \simeq 0.7$
are favored. It is amusing that these values are consistent
with Eq.(\ref{flat}) but the data strongly disfavor
the values $\Omega_m = 1$ of Eq.(\ref{flatflat}). 

Note that a positive $\Omega_{\Lambda}$ acts 
like a negative pressure
which accelerates expansion - 
a normal positive pressure implies that one does work
or adds energy to decrease the volume and increase the pressure:
a positive cosmological constant implies, on the other
hand, that
increase of volume goes with increase of energy,
only possible if the pressure is negative.

\section{QUINTESSENCE.}

The non-zero value $\Omega_{\Lambda} \simeq 0.7$ has two major problems:

\begin{itemize}

\item Its value $(1/100~~eV)^4$ is unnaturally small.
\item At present $\Omega_m$ and $\Omega_{\Lambda}$ are 
the same order of magnitude implying that we live in a 
special era.

\end{itemize}

Both are addressed by {\it quintessence}, an inflaton
field $\Phi$ taylored so that $T_{\mu\nu}(\Phi)
= \Lambda(t)g_{\mu\nu}$. The potential $V(\Phi)$
may be
\begin{equation}
V(\Phi) = M^{4 + \alpha} \Phi^{- \alpha}
\end{equation}
or
\begin{equation}
V(\Phi) = M^4 ( exp(M/\Phi) - 1)
\end{equation}
where $M$ is a parameter [\cite{PR}].

By arranging that $\rho_{\Phi}$ is a little below
$\rho_{\gamma}$ at the end of inflation, it can track
$\rho_{\gamma}$ and then (after matter domination)
$\rho_m$ such that
$\Omega_{\Lambda(t_0)} \sim \Omega_m$ is claimed [\cite{ZWS}]
not to require fine-tuning.
The subject is controversial because, by contrast
to [\cite{ZWS}], [\cite{KL}] claim that slow-roll
inflation and quintessence require fine-tuning
at the level of $1 in 10^{50}$.

More generally, it is well worth examining equations of state
that differ from the one ($\omega = p/\rho = -1$)
implied by constant $\Lambda$. Quintessence covers the possibilities 
$-1 < \omega \leq 0$.

\section{SUMMARY.}

Clearly more data are needed for both the CBR Doppler
peak and the high-Z supernovae. Fortunately both
are expected in the forseeable future.

The current analyses favor
$\Omega_{\Lambda} \simeq 0.7$ and $\Omega_m \simeq 0.3$.

Of course, $\Lambda$ is still 120 orders
of magnitude below its natural value, and 52 orders of 
magnitude below $(250~~GeV)^4$ and that 
theoretical issue remains.

The non-zero $\Lambda$ implies that we live in a special cosmic era: $\Lambda$
was negligible in the past but will
dominate the future
giving exponential growth
$R \sim e^{\Lambda t}, t\rightarrow \infty$.
This cosmic coincidence is addressed by quintessence.

The principal point of our own work in [\cite{FNR}] is that
the value of $l_1$ depends almost completely only on the
geometry of geodesics since recombination, and little
on the details of the accoustic waves, since our iso-$l_1$ plot
agrees well with the numerical results 
of White {\it et al.} [\cite{white}].

\section*{ACKNOWLEDGEMENT}

This work was supported in part by the US Department of Energy
under Grant No. DE-FG05-85ER41036.

\end{document}